\documentclass[
reprint,
superscriptaddress,
amsmath,amssymb,
aps,
]{revtex4-1}

\usepackage{color}
\usepackage{graphicx}% Include figure files
\usepackage[FIGTOPCAP]{subfigure}
\usepackage{dcolumn}% Align table columns on decimal point
\usepackage{bm,braket}% bold math
\begin {document}
%section {title}
%\preprint{APS/123-QED}

\title{Non-self-averaging of current in a totally asymmetric simple exclusion process\\with quenched disorder}% Force line breaks with \\

\author{Issei Sakai}
% \email{6222514@ed.tus.ac.jp}
\affiliation{%
  Department of Physics, Tokyo University of Science, Noda, Chiba 278-8510, Japan
}%

% \author{Keiji Saito}
% \affiliation{%
%   Department of Physics, Keio University, Yokohama, 223-8522, Japan
% }%

\author{Takuma Akimoto}
\email{takuma@rs.tus.ac.jp}
\affiliation{%
  Department of Physics, Tokyo University of Science, Noda, Chiba 278-8510, Japan
}%

%}


\date{\today}% It is always \today, today,

\begin{abstract}
    We investigate the current properties in the totally asymmetric simple exclusion process (TASEP) on a quenched random energy landscape.
    In low- and high-density regimes, the properties are characterized by single-particle dynamics.
    In the intermediate one, the current becomes constant and is maximized.
    Based on the renewal theory, we derive accurate results for the maximum current.
    The maximum current significantly depends on a disorder realization, i.e., non-self-averaging (SA).
    We demonstrate that the disorder average of the maximum current decreases with the system size,
    and the sample-to-sample fluctuations of the maximum current exceed those of current in the low- and high-density regimes.
    We find a significant difference between single-particle dynamics and the TASEP.
    In particular, the non-SA behavior of the maximum current is always observed,
    whereas the transition from non-SA to SA for current in single-particle dynamics exists.
\end{abstract}

%\pacs{05.45.Ac, 05.40.Fb, 87.15.Vv}% PACS, the Physics and Astronomy
\maketitle

%\tableofcontents

%introduction
A heterogeneous environment is one of the origins that generate anomalous diffusion
\cite{PhysRevB.12.2455,BouchaudGeorfes,bardou2002levy,H_fling_2013,PhysRevX.5.011021,PhysRevLett.127.140605,*doi:10.1063/5.0076552,PhysRevE.105.064126}.
Such an environment is often realized by a random energy landscape.
The energy landscape can be divided into two types.
The first type is an annealed energy landscape, where the energy landscape changes with time.
The continuous-time random walk is a diffusion model on an annealed energy landscape \cite{METZLER20001}.
The second type is a quenched energy landscape, where the energy landscape does not change with time.
A typical diffusion model on a quenched random energy landscape is a quenched trap model (QTM) \cite{BouchaudGeorfes}.
Quenched heterogeneous environments are characterized by disorder realizations.
Therefore, for diffusion in quenched heterogeneous environments, the sample-to-sample fluctuations of
the diffusion coefficient  \cite{AkimotoBarkaiSaito,*AkimotoBarkaiSaito2018,LuoYi,AkimotoSaito2020},
mobility \cite{AkimotoSaito2020}, and mean first passage time \cite{AkimotoSaito2019} are essential.
Moreover, these observables become non-self-averaging (SA) \cite{BouchaudGeorfes}.

A pedagogical diffusion model with the many-body effect is the asymmetric simple exclusion process (ASEP) \cite{Derrida}, 
in which hard-core particles diffuse on a one-dimensional lattice.
The ASEP has been applied to various non-equilibrium phenomena, e.g., protein synthesis by ribosomes \cite{ChouLakatou,CiandriniStansfieldRomano,DanaTuller}
and traffic flow \cite{AritaFoulaadvandSanten}.
It belongs to the Kardar-Parisi-Zhang (KPZ) universality class \cite{KardarParisiZhang} 
and is mapped to an interface growth model \cite{KardarParisiZhang,Johansson:2000tk,Tracy:2009tx,Aggarwal}.
In Refs.~\cite{PhysRevLett.104.230602,*SASAMOTO2010523,AmirGideon}, the exact solution to the one-dimensional KPZ equation
was obtained by the weak asymmetric limit of the ASEP.
Moreover, the large deviation function is investigated in the ASEP \cite{PhysRevLett.80.209,PhysRevLett.94.030601} and
symmetric simple exclusion process \cite{PhysRevLett.118.160601}.

The effects of disorder in the ASEP have been extensively studied.
For instance, in the ASEP with heterogeneous hopping rates, the current-density relation exhibits a flat regime for periodic boundary conditions
\cite{TripathyBarma,HarrisStinchcombe,JuhaszSantenIgloi,StinchcombeQueiroz,Nossan_2013,10.1214/14-BJPS277,BanerjeeBasu},
and the first-order phase transition point depends on the disorder for open boundary conditions \cite{Enaud_2004}.
In the ASEP on networks, the current-density relation also exhibits a flat regime \cite{PhysRevLett.107.068702,Neri_2013,PhysRevE.92.052714}.
Moreover, in the non-Poissonian ASEP, which is the ASEP on an annealed random energy landscape, the current becomes freezing \cite{ConcannonBlythe}.
While several many-body effects have been unveiled, the ASEP on a quenched random energy landscape
and the sample-to-sample fluctuations of observables have never been investigated.
Such a heterogeneous system is realized experimentally and considered to be important.
A ribosome diffuses on messenger RNA while decoding the codon and synthesizing protein.
The decoding time becomes heterogeneous as transfer RNA concentration is heterogeneous \cite{DanaTuller}.
In other words, the hopping rate depends on the site, i.e.,
ribosomes diffuse in the quenched random environment.
Moreover, such a heterogeneous environment is also relevant to other transport systems, e.g., proteins on DNA \cite{GraneliGreeneRobertsonYeykal,AustinCoxWang} 
and water transportation in aquaporin \cite{AkimotoHiraoYamamotoYasuiYasuoka}.

In this letter, we discuss how the many-body effect affects SA properties in the ASEP on a quenched energy landscape.
In particular, we demonstrate the SA property of current
and that the fluctuations of the maximum current exceed those of the current in dilute particles or holes.
Therefore, the current fluctuations increase owing to the many-body effect.
When particles and holes are dilute, a transition point from non-SA to SA exists.
However, when the current is maximum, the transition point disappears, and the current is always non-SA.
In Ref.~\cite{sakai2023sample}, we also discuss the SA properties of the diffusion coefficient.

%model
We consider a totally ASEP (TASEP) on a one-dimensional random energy landscape, where the energy landscape is quenched.
It comprises $N$ particles on the lattice of $L$ sites with periodic boundary conditions.
Each site can hold at most one particle.
Quenched disorder means that when realizing the random energy landscape it does not change with time.
At each lattice point, the depth $E>0$ of an energy trap is randomly assigned.
The depths are independent and identically distributed (IID) random variables with an exponential distribution,
$\phi(E)=T_g^{-1}\exp(-E/T_g)$, where $T_g$ is called the glass temperature.
A particle can escape from a trap.
The escape time from a trap is an IID random variable with an exponential distribution.
The mean escape time follows the Arrhenius law, i.e., $\tau_k=\tau_c\exp(E_k/T)$, where $E_k$ is the depth of the energy at site $k$,
$T$ the temperature, and $\tau_c$ a typical time.
Using $\phi(E)$ and the Arrhenius law, it can be proved that the probability density function (PDF) $\psi_\alpha(\tau)$ of 
waiting times follows the power-law distribution:
\begin{equation}
    \int_\tau^\infty d\tau'\psi_\alpha(\tau')\cong\left(\frac{\tau}{\tau_c}\right)^{-\alpha}\ (\tau\geq \tau_c)
    \label{tau_dis}
\end{equation}
with $\alpha\equiv T/T_g$ \cite{AkimotoBarkaiSaito,AkimotoBarkaiSaito2018,bardou2002levy}.

The dynamics of particles are described by the Markovian one in the sense that the waiting time is memory-less. In particular, 
the waiting times at site $k$ are IID random variables following an exponential distribution,
$\psi_k(t_i)=\tau_k^{-1}\exp{(-t_i/\tau_k)}$.
After the waiting time elapses, the particle attempts to hop to the right site. The hop is accepted only if the right site is empty.
When the attempt is a success or failure, the particle is assigned a new waiting time from $\psi_{k+1}(t_i)$ or $\psi_k(t_i)$, respectively.

\begin{figure}[t]
    \centering
    \subfigure[\hspace{8cm}]{\includegraphics[width = 8.6cm]{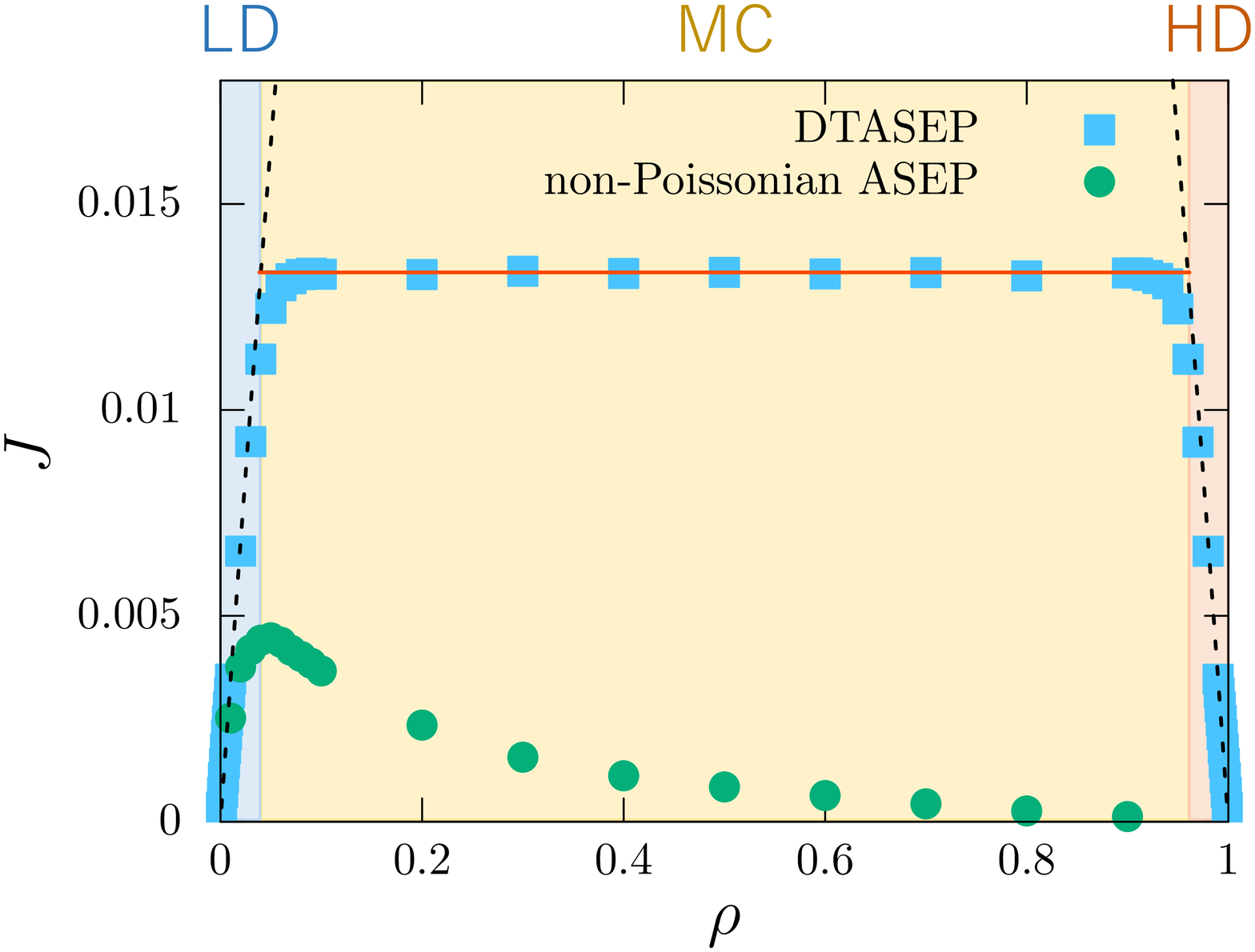}\label{fig:J}}\\
    \subfigure[\hspace{8cm}]{\includegraphics[width = 8.6cm]{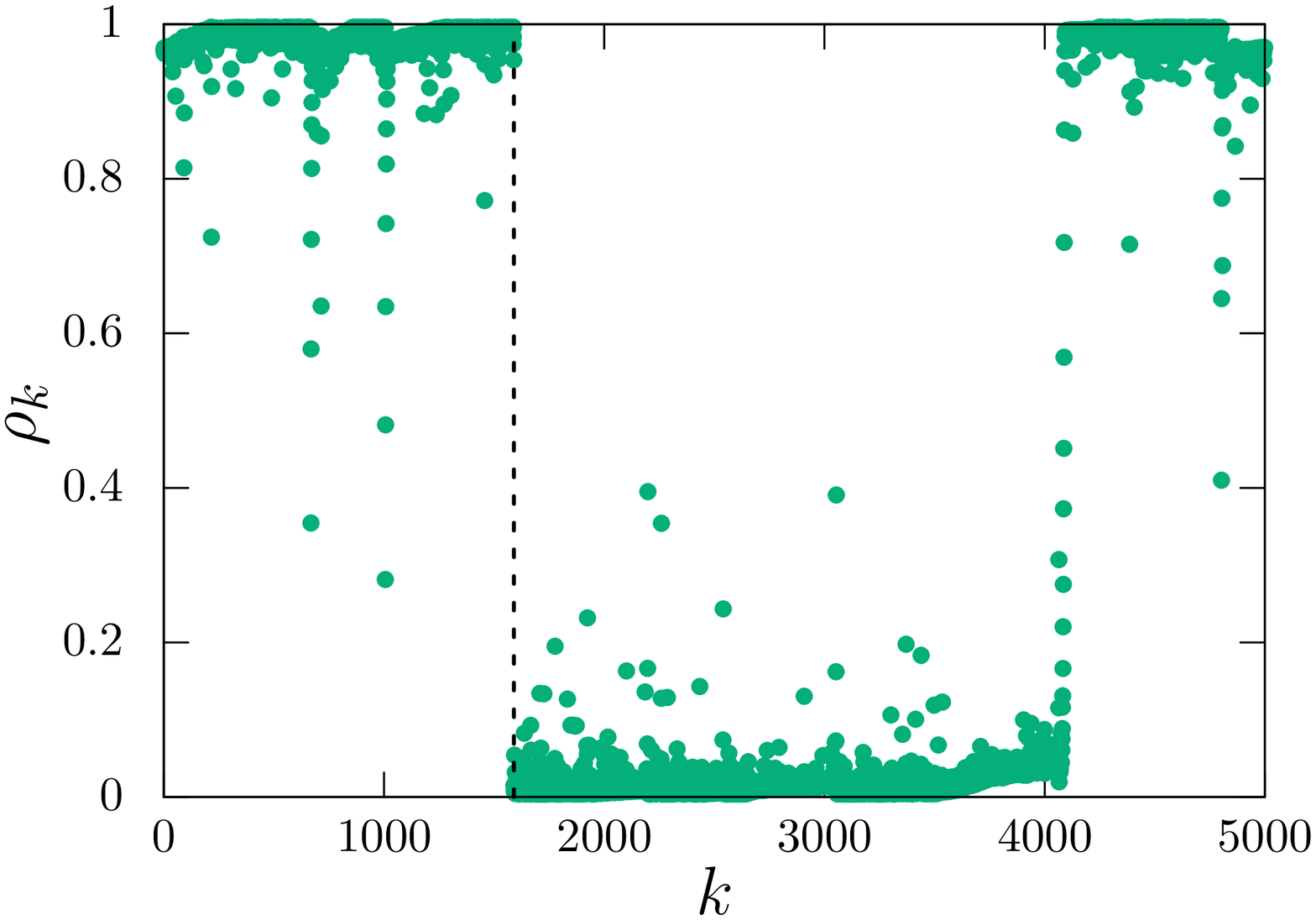}\label{fig:rho}}
    \caption{(a) Current-density relation.
    The squares and the circles are the results of the numerical simulations of the dynamics of the DTASEP ($L=1000$ and $\alpha=1.5$)
    and non-Poissonian TASEP ($L=1000$ and $\alpha=1.5$), respectively.
    The solid line represents Eq.~(\ref{J_max}).
    The dashed line represents Eq.~(\ref{J_TASEP}) for the homogeneous TASEP with $\mu$ being equal to 
    the sample average of the waiting times of the DTASEP.
    (b) Density profile for $\rho=0.5$.
    The dashed line denotes the site with the deepest energy trap.
    The circles are the results of the numerical simulations of the dynamics of the DTASEP ($L=5000$ and $\alpha=1.5$).}
    \label{fig:1}
\end{figure}
%numerical results
The system eventually reaches a steady-state, where the mean current is constant.
Figure~\ref{fig:1}\subref{fig:J} shows the steady-state current $J$ against the particle density $\rho=N/L$,
i.e., the current-density relation, for a disordered TASEP (DTASEP).
For a homogeneous TASEP, the current-density relation is given by \cite{Derrida}
\begin{equation}
    J=\frac{1}{\mu}\rho(1-\rho),
    \label{J_TASEP}
\end{equation}
where $\mu$ is the mean waiting time.
As shown in Fig.~\ref{fig:1}\subref{fig:J}, for the low and high densities, 
the current-density relation coincides with that of the homogeneous TASEP.
For the intermediate regime, the current for the DTASEP deviates from that for the homogeneous TASEP and becomes flat.
Such a flat regime in the DTASEP is also observed in other disorder systems \cite{TripathyBarma,HarrisStinchcombe,JuhaszSantenIgloi,StinchcombeQueiroz,Nossan_2013,10.1214/14-BJPS277,BanerjeeBasu}.
In this regime, the current is independent of particle density and maximized.
Hereafter, we categorize the overall regime into the low density (LD) ($0<\rho\leq \rho^*$), maximum current (MC) ($\rho^*<\rho<1-\rho^*$),
and high density (HD) ($1-\rho^*\leq \rho<1$) regimes (Fig.~\ref{fig:1}\subref{fig:J}).
We explicitly derive the density $\rho^*$ later (see Eq.~(\ref{rho_max})).

Here, we consider the current-density relation of the non-Poissonian TASEP
\cite{ConcannonBlythe} to clarify the effects of a quenched disorder.
The non-Poissonian TASEP is an annealed model. The waiting times do not depend on a site
but are IID random variables.
Moreover, the waiting time distribution follows the power-law distribution.
Figure~\ref{fig:1}\subref{fig:J} shows the current-density relation of the non-Poissonian TASEP when the variance of the waiting time diverges.
The current-density relation is not symmetric and differs from that for the DTASEP.
These discrepancies originate from the condensation front.
For the non-Poissonian ASEP, any site can be the condensation front because the waiting-time distributions at all the sites are identical.
For the DTASEP, only the site with the maximum waiting time can be the condensation front,
at which the segregation of the density profile occurs (Fig.~\ref{fig:1}\subref{fig:rho}).

%MFT
Here, we derive the mean current by the mean-field approximation.
Let $J_k$ be the mean current across the bond between sites $k$ and $k+1$.
In the DTASEP, a hop occurs at a rate $1/\tau_k$.
Thus, the mean current is defined by
\begin{equation}
    J_k=\Braket{\frac{1}{\tau_k}n_k(1-n_{k+1})},
    \label{J_k}
\end{equation}
where $\braket{\cdot}$ is the ensemble average for a fixed disorder and $n_k$ denotes the number of particles at site $k$.
The probability of finding a particle at site $k$ is given by $\rho_k=\braket{n_k}$.
In the mean-field approximation, the correlations between $n_k$ and $n_{k+1}$ can be ignored, implying $\braket{n_kn_{k+1}}\sim\braket{n_k}\braket{n_{k+1}}$.
In the steady-state, the site density is time-independent.
Hence, based on the continuity equation, the current is independent of $k$, i.e., $J_k=J$ for all $k$.
Therefore, we obtain
\begin{equation}
    J=\frac{1}{\tau_k}\rho_k(1-\rho_{k+1}).
    \label{J}
\end{equation}
We note that the right-hand side of Eq.~(\ref{J}) is independent of $k$.

In the low-density limit, $\rho_k\rho_{k+1}$ can be ignored.
Based on the conservation of particles, $\sum_i\rho_i=N$, we obtain
\begin{equation}
    \rho_k\cong\frac{\tau_k}{\bar{\tau}}\rho,
    \label{rho_LD}
\end{equation}
where $\bar{\tau}$ is the sample mean trapping time, $\bar{\tau}\equiv \sum_i\tau_i/L$.
This density profile is the same as the steady-state solution of the master equation of the QTM \cite{AkimotoSaito2020}.
For the HD regime, the particle density is high.
Using the hole density, $\sigma_k=1-\rho_k$; instead of $\rho_k$, we can derive the site density in the same way as in the LD regime.
As a result, we obtain
\begin{equation}
    \rho_k=1-\sigma_k\cong1-\frac{\tau_{k-1}}{\bar{\tau}}(1-\rho).
    \label{rho_HD}
\end{equation}
Multiplying Eq.~(\ref{J}) by $\tau_k/L$ and summing over $k$, we have
\begin{equation}
    \bar{\tau}J=\frac{1}{L}\sum_{k=1}^L\rho_k(1-\rho_{k+1}).
    \label{tauJ}
\end{equation}
In the LD regime, we can ignore $\rho_k\rho_{k+1}$, i.e., the right hand side becomes $\sum_k\rho_k/L=\rho$.
In the HD regime, the right hand side becomes $1-\rho$.
Therefore, Eq.~(\ref{tauJ}) is represented by
\begin{equation}
    \bar{\tau}J\sim
    \left\{\
    \begin{aligned}
        &\rho && (LD\ regime)\\
        &1-\rho && (HD\ regime).
    \end{aligned}
    \right.
    \label{tauJ2}
\end{equation}

To estimate the boundary between the LD and MC regimes in the current-density relation,
we use the current-density relation in the LD regime and the maximum current.
In particular, we define the boundary density $\rho^*$ as the point at which the maximum current $J_{\max}$
and the current-density relation in the LD regime, i.e., $\bar{\tau}^{-1}\rho(1-\rho)$, coincide:
\begin{equation}
    J_{\max}=\frac{1}{\bar{\tau}}\rho^*(1-\rho^*).
    \label{J_ax_rela}
\end{equation}
The current in the MC regime does not strongly depend on the density $\rho$ and is almost equal to $J_{\max}$.
Solving Eq.~(\ref{J_ax_rela}), we obtain the boundary density $\rho^*$,
\begin{equation}
    \rho^*=\frac{1-\sqrt{1-4\bar{\tau}J_{\max}}}{2}.
\end{equation}
For the large-$L$ limit, $J_{\max}$ is much smaller than the current for the homogeneous TASEP, i.e.,
$J_{\max}\ll 1/(4\bar{\tau})$.
Therefore, the boundary density $\rho^*$ can be approximated as
\begin{equation}
    \rho^*\sim \bar{\tau}J_{\max}.
    \label{rho_Jmax}
\end{equation}

We derive the site density in the MC regime.
Since the current does not depend on the site, we have
\begin{equation}
    J_{\max}=\frac{1}{\tau_k}\rho_k(1-\rho_{k+1}).
    \label{J_max_site}
\end{equation}
We assume that both site $k$ and $k+1$ are in the LD phase.
Using Eq.~(\ref{rho_LD}), the site density in the LD phase is given by $\rho_k\cong\tau_k\rho_{LD}/\bar{\tau}$, 
where $\rho_{LD}$ is the particle density in the LD phase.
In the LD phase, $\rho_k\rho_{k+1}$ can be ignored.
It follows that the current is given by $J_{\max}\sim \rho_{LD}/\bar{\tau}$, i.e., $\rho_{LD}\sim\rho^*$.
Therefore, the site density in the LD phase is replaced by $\rho_k\sim\tau_k\rho^*/\bar{\tau}$.
We can derive the site density in the HD phase in the same way as in the LD phase.
As a result, we obtain $\rho_k\sim 1-\tau_{k-1}\rho^*/\bar{\tau}$.

We derive the maximum current based on the segregation of the density profile for the MC regime (Fig.~\ref{fig:1}\subref{fig:rho}).
When the mean waiting time is maximized at site $m$, site $m$ is the boundary between the LD and HD phases.
Therefore, the site density in site $m$ and $m+1$ is represented by $\rho_m\sim 1-\tau_{m-1}\rho^*/\bar{\tau}$ 
and $\rho_{m+1}\sim\tau_{m+1}\rho^*/\bar{\tau}$, respectively.
Using these values and Eq.~(\ref{rho_Jmax}), Eq.~(\ref{J_max_site}) is represented by
\begin{equation}
    J_{\max}\sim\frac{1}{\tau_m}\left(1-\tau_{m-1}J_{\max}\right)\left(1-\tau_{m+1}J_{\max}\right).
\end{equation}
Ignoring the quadratic term of $J_{\max}$ and solving this equation, the maximum current is given by
\begin{equation}
    J_{\max}\sim\frac{1}{\tau_{m-1}+\tau_m+\tau_{m+1}}.
    \label{MC}
\end{equation}
Hereafter, we assume that the mean waiting time is maximized at site $m$.
For $L\rightarrow\infty$, the contribution of $\tau_m$ is stronger than $\tau_{m-1}$ and $\tau_{m+1}$ in Eq.~(\ref{MC}), i.e., $J_{\max}\sim\tau_m^{-1}$.
This result coincides with the previous studies \cite{TripathyBarma,HarrisStinchcombe,StinchcombeQueiroz,Nossan_2013,10.1214/14-BJPS277,BanerjeeBasu}.
Therefore, the boundary density $\rho^*$ can be represented by
\begin{equation}
    \rho^*\sim\frac{\bar{\tau}}{\tau_m}.
    \label{rho_max}
\end{equation}
The scaling of $\tau_m$ follows $\tau_m=O(L^{1/\alpha})$ for $L\rightarrow\infty$.
For $\alpha>1$, we have $\bar{\tau}\rightarrow\braket{\tau}\equiv \int_0^\infty\tau\psi_\alpha(\tau)d\tau$
($L\rightarrow\infty$) by the law of large numbers.
Hence, the scaling of $\rho^*$ becomes
\begin{equation}
    \rho^*\propto L^{-1/\alpha}
\end{equation}
for $\alpha>1$.
The scaling of the sum of $\tau_i$ follows $\sum_i\tau_i=O(L^{1/\alpha})$ for $L\rightarrow\infty$ and $\alpha\leq 1$.
It follows that the scaling of $\rho^*$ becomes
\begin{equation}
    \rho^*\sim L^{-1}\frac{\sum_i\tau_i}{\tau_m}\propto L^{-1}
    \label{rho^*_L}
\end{equation}
for $\alpha\leq 1$.
Therefore, the $\rho^*$ decreases with the system size.

%current
Here, we derive the maximum current by the renewal theory.
We consider the passing of a particle between sites $m$ and $m+1$ as a renewal event.
We call the inter-event time the passage time.
We note that the passage time differs from the first passage time.
The mean of the passage time is given by
\begin{equation}
    \braket{T_m}=\tau_m+\frac{\tau_{m-1}}{\rho_{m-1}}+\frac{\frac{\rho_{m-1}}{\tau_{m-1}}}{\frac{\rho_{m-1}}{\tau_{m-1}}+\frac{1-\rho_{m+2}}{\tau_{m+1}}}\frac{\tau_{m+1}}{1-\rho_{m+2}},
    \label{tau_ave}
\end{equation}
where $\rho_{m-1}\sim1-\tau_{m-2}\rho^*/\bar{\tau}$ and $\rho_{m+2}\sim\tau_{m+2}\rho^*/\bar{\tau}$.
The derivation is given in the Supplemental Material \cite{supplementary}.
We define $n(t)$ as the number of particles passing between sites $m$ and $m+1$ until time $t$.
For the LD and HD regimes, the density profile is homogeneous on a macroscopic scale.
However, the configuration of particles coexists with dilute and dense areas on a microscopic scale at some instant.
When particles are dense on the left of a target site and dilute on the right, the passage time becomes short.
In the opposite case, the passage time becomes long.
Therefore, the passage time depends on the configuration of particles.
For the MC regime, macroscopic density segregation exists (Fig.~\ref{fig:1}\subref{fig:rho}).
Particles are constantly dense on the left of site $m$ and dilute on the right,
but not vice versa.
Therefore, the passage time does not depend on the configuration of particles, i.e., it is almost independent.
Thus, the process of $n(t)$ can be described by the renewal theory \cite{Godreche}.
Based on the renewal theory \cite{Godreche}, the mean number of passing particles is given by $\braket{n(t)}\sim t/\braket{T_m}$ for $t\rightarrow\infty$.
The current is represented by $J=\braket{n(t)}/t$.
Therefore, the maximum current is given by
\begin{equation}
    J_{\mathrm{max}}\sim\frac{1}{\braket{T_m}}\ \ (t\rightarrow\infty).
    \label{J_max}
\end{equation}
The maximum current depends on a disorder realization.
The theory coincides with the numerical simulation (Fig.~\ref{fig:1}\subref{fig:J}).

We consider the effect of disorder on current.
The currents in the LD and HD regimes are given by Eq.~(\ref{tauJ2}), i.e.,
$J\sim\bar{\tau}^{-1}\rho$ and $J\sim\bar{\tau}^{-1}(1-\rho)$, respectively.
When the mean trapping time $\braket{\tau}\equiv \int_0^\infty\tau\psi_\alpha(\tau)d\tau$ is finite ($\alpha>1$), 
we have $\bar{\tau}\rightarrow\braket{\tau}$ ($L\rightarrow\infty$) by the law of large numbers.
In the large-$L$ limit, the current does not depend on the disorder realization.
Therefore, the current is SA \cite{BouchaudGeorfes}.
When the mean trapping time diverges ($\alpha\leq 1$), the law of large numbers breaks down,
but the generalized central limit theorem can be applied to the sum of the mean waiting time.
It states that the PDF of the normalized sum of $\tau_i$ converges to the one-side L\'evy distribution \cite{Feller1971}:
\begin{equation}
    \frac{\sum_{i=1}^L\tau_i}{L^{1/\alpha}}\Rightarrow X_\alpha\ (L\rightarrow\infty),
\end{equation}
where $X_\alpha$ is a random variable following the L\'evy distribution of index $\alpha$.
The currents are given by $J\sim\rho X_\alpha^{-1}L^{1-1/\alpha}$ in the LD regime and $J\sim(1-\rho)X_\alpha^{-1}L^{1-1/\alpha}$
in the HD regime.
Thus, the PDF of $J$ is described by the inverse L\'evy distribution.
Using the first moment of the inverse L\'evy distribution \cite{AkimotoBarkaiSaito}, we obtain the exact asymptotic behavior
of the disorder average of the current:
\begin{equation}
    \braket{J(L)}_{\mathrm{dis}}\sim
    \left\{\
    \begin{aligned}
        &\frac{\rho L^{1-1/\alpha}\Gamma(\alpha^{-1})}{\alpha\tau_c\Gamma(1-\alpha)^{1/\alpha}} && (LD\ regime)\\
        &\frac{(1-\rho)L^{1-1/\alpha}\Gamma(\alpha^{-1})}{\alpha\tau_c\Gamma(1-\alpha)^{1/\alpha}} && (HD\ regime),
    \end{aligned}
    \right.
\end{equation}
where $\braket{\cdot}_{\mathrm{dis}}$ is the disorder averaging, i.e.,
the average obtained under different disorder realizations.
As a result, the current decreases with the system size $L$.

\begin{figure}[t]
    \centering
    \includegraphics[width=8.6cm]{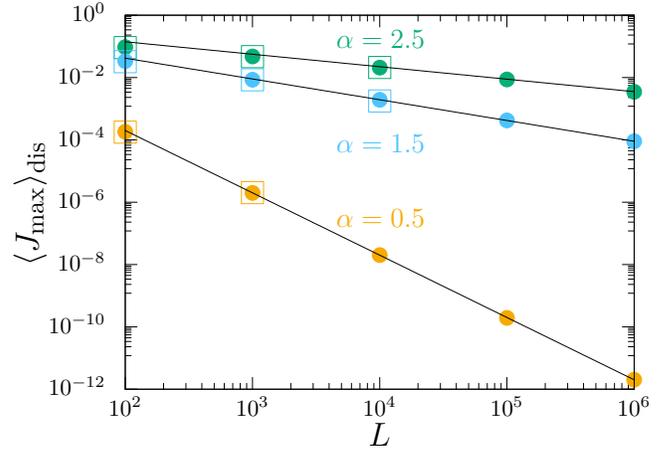}
    \caption{Disorder average of the maximum current as a function of $L$ for several $\alpha$.
    The solid lines are the theoretical curves [Eq.~(\ref{J_dis})].
    The circles are the results of numerical simulations, based on which we calculated the maximum currents (Eq.~(\ref{J_max}))
    for different disorder realizations using Monte Carlo simulations.
    We used $10^4$ disorder realizations.
    The squares are the results of the numerical simulation of the dynamics of the DTASEP.
    We used $10^3$ for $L=10^4$ and $10^4$ disorder realizations for other cases.}
    \label{fig:J_av}
\end{figure}

Next, we consider the effect of disorder on the maximum current.
When the system size is increased, we can find longer and longer $\tau_m$.
Hence, in the large-$L$ limit, we can approximate the passage time as $\braket{T_m}\sim\tau_m$.
In other words, the maximum current is approximated as $J_{\mathrm{max}}\sim\tau_m^{-1}$.
Based on the extreme value theory \cite{HaanFerreira}, we obtain the asymptotic distribution of the maximum current.
As the PDF of the waiting time follows the power-law distribution (Eq.~(\ref{tau_dis})),
the PDF of the normalized $\tau_m$ follows the Fr\'echet distribution \cite{HaanFerreira}:
\begin{equation}
    \frac{\tau_m}{\tau_cL^{1/\alpha}}\Rightarrow Y_\alpha\ (L\rightarrow\infty),
    \label{J_dist}
\end{equation}
where $Y_\alpha$ is a random variable following the Fr\'echet distribution of index $\alpha$.
The PDF of $Y_\alpha$, denoted by $f_\alpha(y)$ with $y>0$ is given by \cite{HaanFerreira}
\begin{equation}
    f_\alpha(y)=\alpha y^{-\alpha-1}\exp{(-y^{-\alpha})}.
    \label{J_dist2}
\end{equation}
As the maximum current is given by $J_{\max}\sim Y_\alpha^{-1}/(\tau_cL^{1/\alpha})$,
the PDF of $J_{\max}$ is described by the inverse Fr\'echet distribution, i.e., the Weibull distribution.
Using the first moment of the Weibull distribution, 
we obtain the exact asymptotic behavior of the disorder average of the maximum current:
\begin{equation}
    \braket{J_{\max}(L)}_{\mathrm{dis}}\sim\frac{1}{\tau_cL^{1/\alpha}}\Gamma\left(1+\frac{1}{\alpha}\right).
    \label{J_dis}
\end{equation}
The maximum current decreases with the system size $L$ for any $\alpha$ (Fig.~\ref{fig:J_av}),
which is different from the result of the QTM. Therefore, this is a manifestation of many-body effects.

\begin{figure}[t]
    \centering
    \includegraphics[width=8.6cm]{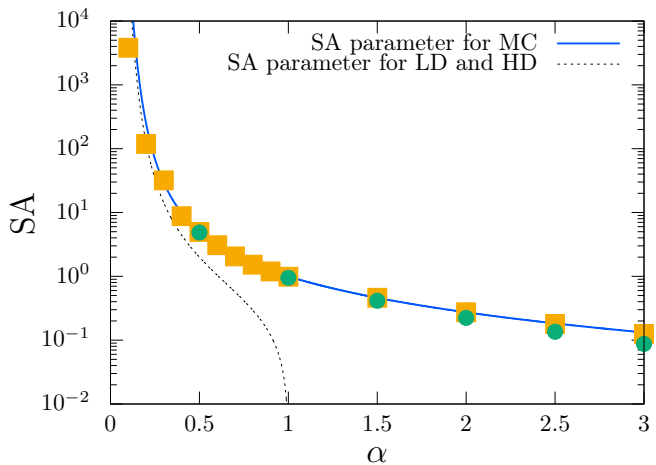}
    \caption{SA parameter of the current as a function of $\alpha$.
    The squares are the results of numerical simulations, where the maximum currents are calculated by Eq.~(\ref{J_max_SA1})
    for different disorder realizations using Monte Carlo simulations ($L=10^5$).
    The circles are obtained using the numerical simulation of the dynamics of the DTASEP ($L=1000$ and $N=500$).
    We used $10^4$ disorder realizations.
    The solid and the dotted lines represent Eqs.~(\ref{J_max_SA2}) and (\ref{J_SA3}), respectively.}
    \label{fig:J_SA}
\end{figure}

%self-averaging
To quantify the SA property, we consider the SA parameter defined as \cite{AkimotoBarkaiSaito} 
\begin{equation}
    \mathrm{SA}(L;J)\equiv \frac{\braket{J(L)^2}_{\mathrm{dis}}-\braket{J(L)}_{\mathrm{dis}}^2}{\braket{J(L)}_{\mathrm{dis}}^2}.
    \label{J_SA1}
\end{equation}
If the SA parameter is $0$, the current is SA.
For the LD and HD regimes, using Eq.~(\ref{J_TASEP}) and $\mu=\bar{\tau}$, we obtain
\begin{equation}
    \mathrm{SA}(L;J)=\frac{\braket{1/\bar{\tau}^2}_{\mathrm{dis}}-\braket{1/\bar{\tau}}_{\mathrm{dis}}^2}{\braket{1/\bar{\tau}}_{\mathrm{dis}}^2},
    \label{J_SA2}
\end{equation}
which is the same as the SA parameter for the diffusion coefficient in the QTM \cite{AkimotoBarkaiSaito}.
Using the first and the second moment of $1/\bar{\tau}$ \cite{AkimotoBarkaiSaito},
we obtain the SA parameter for current:
\begin{equation}
    \lim_{L\rightarrow\infty}\mathrm{SA}(L;J)=
    \left\{\
    \begin{aligned}
        &0 && (\alpha>1)\\
        &\frac{\alpha\Gamma(2/\alpha)}{\Gamma{(1/\alpha)}^2}-1 && (\alpha\leq 1).
    \end{aligned}
    \right.
    \label{J_SA3}
\end{equation}
The SA parameter is a nonzero constant for $\alpha<1$, i.e., $J$ becomes non-SA.
Therefore, the transition from SA to non-SA behavior for the LD and HD regimes exists.

For the MC regimes, using Eq.~(\ref{J_max}), the SA parameter for the maximum current is given by
\begin{equation}
    \mathrm{SA}(L;J_{\mathrm{max}})=\frac{\braket{1/\braket{T_m}^2}_\mathrm{dis}-\braket{1/\braket{T_m}}_\mathrm{dis}^2}{\braket{1/\braket{T_m}}_\mathrm{dis}^2}.
    \label{J_max_SA1}
\end{equation}
For $L\rightarrow\infty$, we can use the following approximation: $\braket{T_m}\sim\tau_m$.
Using the first and the second moment of $1/\tau_m$, we obtain the SA parameter for the maximum current:
\begin{equation}
    \lim_{L\rightarrow\infty}\mathrm{SA}(L;J_{\mathrm{max}})=\frac{\Gamma\left(1+2/\alpha\right)}{\Gamma\left(1+1/\alpha\right)^2}-1.
    \label{J_max_SA2}
\end{equation}
The SA parameter is a nonzero constant for all $\alpha$, i.e., $J_{\mathrm{max}}$ is always non-SA (Fig.~\ref{fig:J_SA}).
Unlike the SA parameter for the LD and HD regimes, no transition from SA to non-SA behavior exists.
When the disorder is not strong, i.e., $\alpha$ is large,
the SA parameter deviates from Eq.~(\ref{J_max_SA2}) because of the contribution of the second and third terms of Eq.~(\ref{tau_ave}) for small $L$.

%conclusion
In summary, we demonstrated the non-SA property of the current in the TASEP on a quenched random energy landscape.
For the LD and HD regimes, the mean current is described by the single-particle dynamics and becomes non-SA for $\alpha<1$.
For the MC regime, based on the renewal theory, we derived the exact expression for the maximum current.
In particular, for $L\rightarrow\infty$, the longest waiting time determines the maximum current.
We demonstrated that the PDF of the maximum current follows the Weibull distribution in the large-$L$ limit.
Moreover, we introduced the SA parameter to quantify the non-SA property.
For the LD and HD regimes, the transition point between non-SA and SA is $\alpha=1$.
For the MC regime, the transition point disappears, and the maximum current becomes non-SA for all $\alpha$.
This non-SA behavior for $\alpha>1$ is a manifestation of the many-body effect in transport on a quenched random energy landscape.

\begin{acknowledgments}
    We thank K. Saito for a fruitful discussion.
    T.A. was supported by JSPS Grant-in-Aid for Scientific Research (No. C JP21K033920).
\end{acknowledgments}

\bibliographystyle{apsrev4-1}
\bibliography{ms}

\begin{thebibliography}{49}%
\makeatletter
\providecommand \@ifxundefined [1]{%
 \@ifx{#1\undefined}
}%
\providecommand \@ifnum [1]{%
 \ifnum #1\expandafter \@firstoftwo
 \else \expandafter \@secondoftwo
 \fi
}%
\providecommand \@ifx [1]{%
 \ifx #1\expandafter \@firstoftwo
 \else \expandafter \@secondoftwo
 \fi
}%
\providecommand \natexlab [1]{#1}%
\providecommand \enquote  [1]{``#1''}%
\providecommand \bibnamefont  [1]{#1}%
\providecommand \bibfnamefont [1]{#1}%
\providecommand \citenamefont [1]{#1}%
\providecommand \href@noop [0]{\@secondoftwo}%
\providecommand \href [0]{\begingroup \@sanitize@url \@href}%
\providecommand \@href[1]{\@@startlink{#1}\@@href}%
\providecommand \@@href[1]{\endgroup#1\@@endlink}%
\providecommand \@sanitize@url [0]{\catcode `\\12\catcode `\$12\catcode
  `\&12\catcode `\#12\catcode `\^12\catcode `\_12\catcode `\%12\relax}%
\providecommand \@@startlink[1]{}%
\providecommand \@@endlink[0]{}%
\providecommand \url  [0]{\begingroup\@sanitize@url \@url }%
\providecommand \@url [1]{\endgroup\@href {#1}{\urlprefix }}%
\providecommand \urlprefix  [0]{URL }%
\providecommand \Eprint [0]{\href }%
\providecommand \doibase [0]{http://dx.doi.org/}%
\providecommand \selectlanguage [0]{\@gobble}%
\providecommand \bibinfo  [0]{\@secondoftwo}%
\providecommand \bibfield  [0]{\@secondoftwo}%
\providecommand \translation [1]{[#1]}%
\providecommand \BibitemOpen [0]{}%
\providecommand \bibitemStop [0]{}%
\providecommand \bibitemNoStop [0]{.\EOS\space}%
\providecommand \EOS [0]{\spacefactor3000\relax}%
\providecommand \BibitemShut  [1]{\csname bibitem#1\endcsname}%
\let\auto@bib@innerbib\@empty
%</preamble>
\bibitem [{\citenamefont {Scher}\ and\ \citenamefont
  {Montroll}(1975)}]{PhysRevB.12.2455}%
  \BibitemOpen
  \bibfield  {author} {\bibinfo {author} {\bibfnamefont {H.}~\bibnamefont
  {Scher}}\ and\ \bibinfo {author} {\bibfnamefont {E.~W.}\ \bibnamefont
  {Montroll}},\ }\href {\doibase 10.1103/PhysRevB.12.2455} {\bibfield
  {journal} {\bibinfo  {journal} {Phys. Rev. B}\ }\textbf {\bibinfo {volume}
  {12}},\ \bibinfo {pages} {2455} (\bibinfo {year} {1975})}\BibitemShut
  {NoStop}%
\bibitem [{\citenamefont {Bouchaud}\ and\ \citenamefont
  {Georges}(1990)}]{BouchaudGeorfes}%
  \BibitemOpen
  \bibfield  {author} {\bibinfo {author} {\bibfnamefont {J.~P.}\ \bibnamefont
  {Bouchaud}}\ and\ \bibinfo {author} {\bibfnamefont {A.}~\bibnamefont
  {Georges}},\ }\href@noop {} {\bibfield  {journal} {\bibinfo  {journal} {Phy.
  Rep.}\ }\textbf {\bibinfo {volume} {195}} (\bibinfo {year}
  {1990})}\BibitemShut {NoStop}%
\bibitem [{\citenamefont {Bardou}\ \emph {et~al.}(2002)\citenamefont {Bardou},
  \citenamefont {Bouchaud}, \citenamefont {Aspect},\ and\ \citenamefont
  {Cohen-Tannoudji}}]{bardou2002levy}%
  \BibitemOpen
  \bibfield  {author} {\bibinfo {author} {\bibfnamefont {F.}~\bibnamefont
  {Bardou}}, \bibinfo {author} {\bibfnamefont {J.~P.}\ \bibnamefont
  {Bouchaud}}, \bibinfo {author} {\bibfnamefont {A.}~\bibnamefont {Aspect}}, \
  and\ \bibinfo {author} {\bibfnamefont {C.}~\bibnamefont {Cohen-Tannoudji}},\
  }\href@noop {} {\emph {\bibinfo {title} {L{\'e}vy statistics and laser
  cooling: how rare events bring atoms to rest}}}\ (\bibinfo  {publisher}
  {Cambridge University Press},\ \bibinfo {year} {2002})\BibitemShut {NoStop}%
\bibitem [{\citenamefont {H{\"o}fling}\ and\ \citenamefont
  {Franosch}(2013)}]{H_fling_2013}%
  \BibitemOpen
  \bibfield  {author} {\bibinfo {author} {\bibfnamefont {F.}~\bibnamefont
  {H{\"o}fling}}\ and\ \bibinfo {author} {\bibfnamefont {T.}~\bibnamefont
  {Franosch}},\ }\href {\doibase 10.1088/0034-4885/76/4/046602} {\bibfield
  {journal} {\bibinfo  {journal} {Rep. Prog. Phys.}\ }\textbf {\bibinfo
  {volume} {76}},\ \bibinfo {pages} {046602} (\bibinfo {year}
  {2013})}\BibitemShut {NoStop}%
\bibitem [{\citenamefont {Manzo}\ \emph {et~al.}(2015)\citenamefont {Manzo},
  \citenamefont {Torreno-Pina}, \citenamefont {Massignan}, \citenamefont
  {Lapeyre}, \citenamefont {Lewenstein},\ and\ \citenamefont
  {Garcia~Parajo}}]{PhysRevX.5.011021}%
  \BibitemOpen
  \bibfield  {author} {\bibinfo {author} {\bibfnamefont {C.}~\bibnamefont
  {Manzo}}, \bibinfo {author} {\bibfnamefont {J.~A.}\ \bibnamefont
  {Torreno-Pina}}, \bibinfo {author} {\bibfnamefont {P.}~\bibnamefont
  {Massignan}}, \bibinfo {author} {\bibfnamefont {G.~J.}\ \bibnamefont
  {Lapeyre}}, \bibinfo {author} {\bibfnamefont {M.}~\bibnamefont {Lewenstein}},
  \ and\ \bibinfo {author} {\bibfnamefont {M.~F.}\ \bibnamefont
  {Garcia~Parajo}},\ }\href {\doibase 10.1103/PhysRevX.5.011021} {\bibfield
  {journal} {\bibinfo  {journal} {Phys. Rev. X}\ }\textbf {\bibinfo {volume}
  {5}},\ \bibinfo {pages} {011021} (\bibinfo {year} {2015})}\BibitemShut
  {NoStop}%
\bibitem [{\citenamefont {Barkai}\ \emph {et~al.}(2021)\citenamefont {Barkai},
  \citenamefont {Radons},\ and\ \citenamefont
  {Akimoto}}]{PhysRevLett.127.140605}%
  \BibitemOpen
  \bibfield  {author} {\bibinfo {author} {\bibfnamefont {E.}~\bibnamefont
  {Barkai}}, \bibinfo {author} {\bibfnamefont {G.}~\bibnamefont {Radons}}, \
  and\ \bibinfo {author} {\bibfnamefont {T.}~\bibnamefont {Akimoto}},\ }\href
  {\doibase 10.1103/PhysRevLett.127.140605} {\bibfield  {journal} {\bibinfo
  {journal} {Phys. Rev. Lett.}\ }\textbf {\bibinfo {volume} {127}},\ \bibinfo
  {pages} {140605} (\bibinfo {year} {2021})}\BibitemShut {NoStop}%
\bibitem [{\citenamefont {Barkai}\ \emph {et~al.}(2022)\citenamefont {Barkai},
  \citenamefont {Radons},\ and\ \citenamefont
  {Akimoto}}]{doi:10.1063/5.0076552}%
  \BibitemOpen
  \bibfield  {author} {\bibinfo {author} {\bibfnamefont {E.}~\bibnamefont
  {Barkai}}, \bibinfo {author} {\bibfnamefont {G.}~\bibnamefont {Radons}}, \
  and\ \bibinfo {author} {\bibfnamefont {T.}~\bibnamefont {Akimoto}},\ }\href
  {\doibase 10.1063/5.0076552} {\bibfield  {journal} {\bibinfo  {journal} {J.
  Chem. Phys.}\ }\textbf {\bibinfo {volume} {156}},\ \bibinfo {pages} {044118}
  (\bibinfo {year} {2022})}\BibitemShut {NoStop}%
\bibitem [{\citenamefont {Akimoto}\ \emph {et~al.}(2022)\citenamefont
  {Akimoto}, \citenamefont {Barkai},\ and\ \citenamefont
  {Radons}}]{PhysRevE.105.064126}%
  \BibitemOpen
  \bibfield  {author} {\bibinfo {author} {\bibfnamefont {T.}~\bibnamefont
  {Akimoto}}, \bibinfo {author} {\bibfnamefont {E.}~\bibnamefont {Barkai}}, \
  and\ \bibinfo {author} {\bibfnamefont {G.}~\bibnamefont {Radons}},\ }\href
  {\doibase 10.1103/PhysRevE.105.064126} {\bibfield  {journal} {\bibinfo
  {journal} {Phys. Rev. E}\ }\textbf {\bibinfo {volume} {105}},\ \bibinfo
  {pages} {064126} (\bibinfo {year} {2022})}\BibitemShut {NoStop}%
\bibitem [{\citenamefont {Metzler}\ and\ \citenamefont
  {Klafter}(2000)}]{METZLER20001}%
  \BibitemOpen
  \bibfield  {author} {\bibinfo {author} {\bibfnamefont {R.}~\bibnamefont
  {Metzler}}\ and\ \bibinfo {author} {\bibfnamefont {J.}~\bibnamefont
  {Klafter}},\ }\href {\doibase https://doi.org/10.1016/S0370-1573(00)00070-3}
  {\bibfield  {journal} {\bibinfo  {journal} {Phys. Rep.}\ }\textbf {\bibinfo
  {volume} {339}},\ \bibinfo {pages} {1} (\bibinfo {year} {2000})}\BibitemShut
  {NoStop}%
\bibitem [{\citenamefont {Akimoto}\ \emph {et~al.}(2016)\citenamefont
  {Akimoto}, \citenamefont {Barkai},\ and\ \citenamefont
  {Saito}}]{AkimotoBarkaiSaito}%
  \BibitemOpen
  \bibfield  {author} {\bibinfo {author} {\bibfnamefont {T.}~\bibnamefont
  {Akimoto}}, \bibinfo {author} {\bibfnamefont {E.}~\bibnamefont {Barkai}}, \
  and\ \bibinfo {author} {\bibfnamefont {K.}~\bibnamefont {Saito}},\ }\href
  {\doibase 10.1103/PhysRevLett.117.180602} {\bibfield  {journal} {\bibinfo
  {journal} {Phys. Rev. Lett.}\ }\textbf {\bibinfo {volume} {117}},\ \bibinfo
  {pages} {180602} (\bibinfo {year} {2016})}\BibitemShut {NoStop}%
\bibitem [{\citenamefont {Akimoto}\ \emph {et~al.}(2018)\citenamefont
  {Akimoto}, \citenamefont {Barkai},\ and\ \citenamefont
  {Saito}}]{AkimotoBarkaiSaito2018}%
  \BibitemOpen
  \bibfield  {author} {\bibinfo {author} {\bibfnamefont {T.}~\bibnamefont
  {Akimoto}}, \bibinfo {author} {\bibfnamefont {E.}~\bibnamefont {Barkai}}, \
  and\ \bibinfo {author} {\bibfnamefont {K.}~\bibnamefont {Saito}},\ }\href
  {\doibase 10.1103/PhysRevE.97.052143} {\bibfield  {journal} {\bibinfo
  {journal} {Phys. Rev. E}\ }\textbf {\bibinfo {volume} {97}},\ \bibinfo
  {pages} {052143} (\bibinfo {year} {2018})}\BibitemShut {NoStop}%
\bibitem [{\citenamefont {Luo}\ and\ \citenamefont {Yi}(2018)}]{LuoYi}%
  \BibitemOpen
  \bibfield  {author} {\bibinfo {author} {\bibfnamefont {L.}~\bibnamefont
  {Luo}}\ and\ \bibinfo {author} {\bibfnamefont {M.}~\bibnamefont {Yi}},\
  }\href {\doibase 10.1103/PhysRevE.97.042122} {\bibfield  {journal} {\bibinfo
  {journal} {Phys. Rev. E}\ }\textbf {\bibinfo {volume} {97}},\ \bibinfo
  {pages} {042122} (\bibinfo {year} {2018})}\BibitemShut {NoStop}%
\bibitem [{\citenamefont {Akimoto}\ and\ \citenamefont
  {Saito}(2019)}]{AkimotoSaito2019}%
  \BibitemOpen
  \bibfield  {author} {\bibinfo {author} {\bibfnamefont {T.}~\bibnamefont
  {Akimoto}}\ and\ \bibinfo {author} {\bibfnamefont {K.}~\bibnamefont
  {Saito}},\ }\href {\doibase 10.1103/PhysRevE.99.052127} {\bibfield  {journal}
  {\bibinfo  {journal} {Phys. Rev. E}\ }\textbf {\bibinfo {volume} {99}},\
  \bibinfo {pages} {052127} (\bibinfo {year} {2019})}\BibitemShut {NoStop}%
\bibitem [{\citenamefont {Akimoto}\ and\ \citenamefont
  {Saito}(2020)}]{AkimotoSaito2020}%
  \BibitemOpen
  \bibfield  {author} {\bibinfo {author} {\bibfnamefont {T.}~\bibnamefont
  {Akimoto}}\ and\ \bibinfo {author} {\bibfnamefont {K.}~\bibnamefont
  {Saito}},\ }\href {\doibase 10.1103/PhysRevE.101.042133} {\bibfield
  {journal} {\bibinfo  {journal} {Phys. Rev. E}\ }\textbf {\bibinfo {volume}
  {101}},\ \bibinfo {pages} {042133} (\bibinfo {year} {2020})}\BibitemShut
  {NoStop}%
\bibitem [{\citenamefont {Derrida}(1998)}]{Derrida}%
  \BibitemOpen
  \bibfield  {author} {\bibinfo {author} {\bibfnamefont {B.}~\bibnamefont
  {Derrida}},\ }\href {\doibase https://doi.org/10.1016/S0370-1573(98)00006-4}
  {\bibfield  {journal} {\bibinfo  {journal} {Phys. Rep.}\ }\textbf {\bibinfo
  {volume} {301}},\ \bibinfo {pages} {65} (\bibinfo {year} {1998})}\BibitemShut
  {NoStop}%
\bibitem [{\citenamefont {Chou}\ and\ \citenamefont
  {Lakatos}(2004)}]{ChouLakatou}%
  \BibitemOpen
  \bibfield  {author} {\bibinfo {author} {\bibfnamefont {T.}~\bibnamefont
  {Chou}}\ and\ \bibinfo {author} {\bibfnamefont {G.}~\bibnamefont {Lakatos}},\
  }\href {\doibase 10.1103/PhysRevLett.93.198101} {\bibfield  {journal}
  {\bibinfo  {journal} {Phys. Rev. Lett.}\ }\textbf {\bibinfo {volume} {93}},\
  \bibinfo {pages} {198101} (\bibinfo {year} {2004})}\BibitemShut {NoStop}%
\bibitem [{\citenamefont {Ciandrini}\ \emph {et~al.}(2010)\citenamefont
  {Ciandrini}, \citenamefont {Stansfield},\ and\ \citenamefont
  {Romano}}]{CiandriniStansfieldRomano}%
  \BibitemOpen
  \bibfield  {author} {\bibinfo {author} {\bibfnamefont {L.}~\bibnamefont
  {Ciandrini}}, \bibinfo {author} {\bibfnamefont {I.}~\bibnamefont
  {Stansfield}}, \ and\ \bibinfo {author} {\bibfnamefont {M.~C.}\ \bibnamefont
  {Romano}},\ }\href {\doibase 10.1103/PhysRevE.81.051904} {\bibfield
  {journal} {\bibinfo  {journal} {Phys. Rev. E}\ }\textbf {\bibinfo {volume}
  {81}},\ \bibinfo {pages} {051904} (\bibinfo {year} {2010})}\BibitemShut
  {NoStop}%
\bibitem [{\citenamefont {Dana}\ and\ \citenamefont
  {Tuller}(2014)}]{DanaTuller}%
  \BibitemOpen
  \bibfield  {author} {\bibinfo {author} {\bibfnamefont {A.}~\bibnamefont
  {Dana}}\ and\ \bibinfo {author} {\bibfnamefont {T.}~\bibnamefont {Tuller}},\
  }\href {\doibase 10.1093/nar/gku646} {\bibfield  {journal} {\bibinfo
  {journal} {Nucleic Acids Research}\ }\textbf {\bibinfo {volume} {42}},\
  \bibinfo {pages} {9171} (\bibinfo {year} {2014})}\BibitemShut {NoStop}%
\bibitem [{\citenamefont {Arita}\ \emph {et~al.}(2017)\citenamefont {Arita},
  \citenamefont {Foulaadvand},\ and\ \citenamefont
  {Santen}}]{AritaFoulaadvandSanten}%
  \BibitemOpen
  \bibfield  {author} {\bibinfo {author} {\bibfnamefont {C.}~\bibnamefont
  {Arita}}, \bibinfo {author} {\bibfnamefont {M.~E.}\ \bibnamefont
  {Foulaadvand}}, \ and\ \bibinfo {author} {\bibfnamefont {L.}~\bibnamefont
  {Santen}},\ }\href {\doibase 10.1103/PhysRevE.95.032108} {\bibfield
  {journal} {\bibinfo  {journal} {Phys. Rev. E}\ }\textbf {\bibinfo {volume}
  {95}},\ \bibinfo {pages} {032108} (\bibinfo {year} {2017})}\BibitemShut
  {NoStop}%
\bibitem [{\citenamefont {Kardar}\ \emph {et~al.}(1986)\citenamefont {Kardar},
  \citenamefont {Parisi},\ and\ \citenamefont {Zhang}}]{KardarParisiZhang}%
  \BibitemOpen
  \bibfield  {author} {\bibinfo {author} {\bibfnamefont {M.}~\bibnamefont
  {Kardar}}, \bibinfo {author} {\bibfnamefont {G.}~\bibnamefont {Parisi}}, \
  and\ \bibinfo {author} {\bibfnamefont {Y.-C.}\ \bibnamefont {Zhang}},\ }\href
  {\doibase 10.1103/PhysRevLett.56.889} {\bibfield  {journal} {\bibinfo
  {journal} {Phys. Rev. Lett.}\ }\textbf {\bibinfo {volume} {56}},\ \bibinfo
  {pages} {889} (\bibinfo {year} {1986})}\BibitemShut {NoStop}%
\bibitem [{\citenamefont {Johansson}(2000)}]{Johansson:2000tk}%
  \BibitemOpen
  \bibfield  {author} {\bibinfo {author} {\bibfnamefont {K.}~\bibnamefont
  {Johansson}},\ }\href {\doibase 10.1007/s002200050027} {\bibfield  {journal}
  {\bibinfo  {journal} {Comm. Math. Phys.}\ }\textbf {\bibinfo {volume}
  {209}},\ \bibinfo {pages} {437} (\bibinfo {year} {2000})}\BibitemShut
  {NoStop}%
\bibitem [{\citenamefont {Tracy}\ and\ \citenamefont
  {Widom}(2009)}]{Tracy:2009tx}%
  \BibitemOpen
  \bibfield  {author} {\bibinfo {author} {\bibfnamefont {C.~A.}\ \bibnamefont
  {Tracy}}\ and\ \bibinfo {author} {\bibfnamefont {H.}~\bibnamefont {Widom}},\
  }\href {\doibase 10.1007/s00220-009-0761-0} {\bibfield  {journal} {\bibinfo
  {journal} {Comm. Math. Phys.}\ }\textbf {\bibinfo {volume} {290}},\ \bibinfo
  {pages} {129} (\bibinfo {year} {2009})}\BibitemShut {NoStop}%
\bibitem [{\citenamefont {Aggarwal}(2018)}]{Aggarwal}%
  \BibitemOpen
  \bibfield  {author} {\bibinfo {author} {\bibfnamefont {A.}~\bibnamefont
  {Aggarwal}},\ }\href {\doibase 10.1215/00127094-2017-0029} {\bibfield
  {journal} {\bibinfo  {journal} {Duke Math. J.}\ }\textbf {\bibinfo {volume}
  {167}},\ \bibinfo {pages} {269 } (\bibinfo {year} {2018})}\BibitemShut
  {NoStop}%
\bibitem [{\citenamefont {Sasamoto}\ and\ \citenamefont
  {Spohn}(2010{\natexlab{a}})}]{PhysRevLett.104.230602}%
  \BibitemOpen
  \bibfield  {author} {\bibinfo {author} {\bibfnamefont {T.}~\bibnamefont
  {Sasamoto}}\ and\ \bibinfo {author} {\bibfnamefont {H.}~\bibnamefont
  {Spohn}},\ }\href {\doibase 10.1103/PhysRevLett.104.230602} {\bibfield
  {journal} {\bibinfo  {journal} {Phys. Rev. Lett.}\ }\textbf {\bibinfo
  {volume} {104}},\ \bibinfo {pages} {230602} (\bibinfo {year}
  {2010}{\natexlab{a}})}\BibitemShut {NoStop}%
\bibitem [{\citenamefont {Sasamoto}\ and\ \citenamefont
  {Spohn}(2010{\natexlab{b}})}]{SASAMOTO2010523}%
  \BibitemOpen
  \bibfield  {author} {\bibinfo {author} {\bibfnamefont {T.}~\bibnamefont
  {Sasamoto}}\ and\ \bibinfo {author} {\bibfnamefont {H.}~\bibnamefont
  {Spohn}},\ }\href {\doibase https://doi.org/10.1016/j.nuclphysb.2010.03.026}
  {\bibfield  {journal} {\bibinfo  {journal} {Nuclear Phys. B}\ }\textbf
  {\bibinfo {volume} {834}},\ \bibinfo {pages} {523} (\bibinfo {year}
  {2010}{\natexlab{b}})}\BibitemShut {NoStop}%
\bibitem [{\citenamefont {Amir}\ \emph {et~al.}(2011)\citenamefont {Amir},
  \citenamefont {Corwin},\ and\ \citenamefont {Quastel}}]{AmirGideon}%
  \BibitemOpen
  \bibfield  {author} {\bibinfo {author} {\bibfnamefont {G.}~\bibnamefont
  {Amir}}, \bibinfo {author} {\bibfnamefont {I.}~\bibnamefont {Corwin}}, \ and\
  \bibinfo {author} {\bibfnamefont {J.}~\bibnamefont {Quastel}},\ }\href
  {\doibase https://doi.org/10.1002/cpa.20347} {\bibfield  {journal} {\bibinfo
  {journal} {Comm. Pure Appl. Math.}\ }\textbf {\bibinfo {volume} {64}},\
  \bibinfo {pages} {466} (\bibinfo {year} {2011})}\BibitemShut {NoStop}%
\bibitem [{\citenamefont {Derrida}\ and\ \citenamefont
  {Lebowitz}(1998)}]{PhysRevLett.80.209}%
  \BibitemOpen
  \bibfield  {author} {\bibinfo {author} {\bibfnamefont {B.}~\bibnamefont
  {Derrida}}\ and\ \bibinfo {author} {\bibfnamefont {J.~L.}\ \bibnamefont
  {Lebowitz}},\ }\href {\doibase 10.1103/PhysRevLett.80.209} {\bibfield
  {journal} {\bibinfo  {journal} {Phys. Rev. Lett.}\ }\textbf {\bibinfo
  {volume} {80}},\ \bibinfo {pages} {209} (\bibinfo {year} {1998})}\BibitemShut
  {NoStop}%
\bibitem [{\citenamefont {Bertini}\ \emph {et~al.}(2005)\citenamefont
  {Bertini}, \citenamefont {De~Sole}, \citenamefont {Gabrielli}, \citenamefont
  {Jona-Lasinio},\ and\ \citenamefont {Landim}}]{PhysRevLett.94.030601}%
  \BibitemOpen
  \bibfield  {author} {\bibinfo {author} {\bibfnamefont {L.}~\bibnamefont
  {Bertini}}, \bibinfo {author} {\bibfnamefont {A.}~\bibnamefont {De~Sole}},
  \bibinfo {author} {\bibfnamefont {D.}~\bibnamefont {Gabrielli}}, \bibinfo
  {author} {\bibfnamefont {G.}~\bibnamefont {Jona-Lasinio}}, \ and\ \bibinfo
  {author} {\bibfnamefont {C.}~\bibnamefont {Landim}},\ }\href {\doibase
  10.1103/PhysRevLett.94.030601} {\bibfield  {journal} {\bibinfo  {journal}
  {Phys. Rev. Lett.}\ }\textbf {\bibinfo {volume} {94}},\ \bibinfo {pages}
  {030601} (\bibinfo {year} {2005})}\BibitemShut {NoStop}%
\bibitem [{\citenamefont {Imamura}\ \emph {et~al.}(2017)\citenamefont
  {Imamura}, \citenamefont {Mallick},\ and\ \citenamefont
  {Sasamoto}}]{PhysRevLett.118.160601}%
  \BibitemOpen
  \bibfield  {author} {\bibinfo {author} {\bibfnamefont {T.}~\bibnamefont
  {Imamura}}, \bibinfo {author} {\bibfnamefont {K.}~\bibnamefont {Mallick}}, \
  and\ \bibinfo {author} {\bibfnamefont {T.}~\bibnamefont {Sasamoto}},\ }\href
  {\doibase 10.1103/PhysRevLett.118.160601} {\bibfield  {journal} {\bibinfo
  {journal} {Phys. Rev. Lett.}\ }\textbf {\bibinfo {volume} {118}},\ \bibinfo
  {pages} {160601} (\bibinfo {year} {2017})}\BibitemShut {NoStop}%
\bibitem [{\citenamefont {Tripathy}\ and\ \citenamefont
  {Barma}(1998)}]{TripathyBarma}%
  \BibitemOpen
  \bibfield  {author} {\bibinfo {author} {\bibfnamefont {G.}~\bibnamefont
  {Tripathy}}\ and\ \bibinfo {author} {\bibfnamefont {M.}~\bibnamefont
  {Barma}},\ }\href {\doibase 10.1103/PhysRevE.58.1911} {\bibfield  {journal}
  {\bibinfo  {journal} {Phys. Rev. E}\ }\textbf {\bibinfo {volume} {58}},\
  \bibinfo {pages} {1911} (\bibinfo {year} {1998})}\BibitemShut {NoStop}%
\bibitem [{\citenamefont {Harris}\ and\ \citenamefont
  {Stinchcombe}(2004)}]{HarrisStinchcombe}%
  \BibitemOpen
  \bibfield  {author} {\bibinfo {author} {\bibfnamefont {R.~J.}\ \bibnamefont
  {Harris}}\ and\ \bibinfo {author} {\bibfnamefont {R.~B.}\ \bibnamefont
  {Stinchcombe}},\ }\href {\doibase 10.1103/PhysRevE.70.016108} {\bibfield
  {journal} {\bibinfo  {journal} {Phys. Rev. E}\ }\textbf {\bibinfo {volume}
  {70}},\ \bibinfo {pages} {016108} (\bibinfo {year} {2004})}\BibitemShut
  {NoStop}%
\bibitem [{\citenamefont {Juh\'asz}\ \emph {et~al.}(2006)\citenamefont
  {Juh\'asz}, \citenamefont {Santen},\ and\ \citenamefont
  {Igl\'oi}}]{JuhaszSantenIgloi}%
  \BibitemOpen
  \bibfield  {author} {\bibinfo {author} {\bibfnamefont {R.}~\bibnamefont
  {Juh\'asz}}, \bibinfo {author} {\bibfnamefont {L.}~\bibnamefont {Santen}}, \
  and\ \bibinfo {author} {\bibfnamefont {F.}~\bibnamefont {Igl\'oi}},\ }\href
  {\doibase 10.1103/PhysRevE.74.061101} {\bibfield  {journal} {\bibinfo
  {journal} {Phys. Rev. E}\ }\textbf {\bibinfo {volume} {74}},\ \bibinfo
  {pages} {061101} (\bibinfo {year} {2006})}\BibitemShut {NoStop}%
\bibitem [{\citenamefont {Stinchcombe}\ and\ \citenamefont
  {de~Queiroz}(2011)}]{StinchcombeQueiroz}%
  \BibitemOpen
  \bibfield  {author} {\bibinfo {author} {\bibfnamefont {R.~B.}\ \bibnamefont
  {Stinchcombe}}\ and\ \bibinfo {author} {\bibfnamefont {S.~L.~A.}\
  \bibnamefont {de~Queiroz}},\ }\href {\doibase 10.1103/PhysRevE.83.061113}
  {\bibfield  {journal} {\bibinfo  {journal} {Phys. Rev. E}\ }\textbf {\bibinfo
  {volume} {83}},\ \bibinfo {pages} {061113} (\bibinfo {year}
  {2011})}\BibitemShut {NoStop}%
\bibitem [{\citenamefont {Nossan}(2013)}]{Nossan_2013}%
  \BibitemOpen
  \bibfield  {author} {\bibinfo {author} {\bibfnamefont {J.~S.}\ \bibnamefont
  {Nossan}},\ }\href {\doibase 10.1088/1751-8113/46/31/315001} {\bibfield
  {journal} {\bibinfo  {journal} {Journal of Physics A: Mathematical and
  Theoretical}\ }\textbf {\bibinfo {volume} {46}},\ \bibinfo {pages} {315001}
  (\bibinfo {year} {2013})}\BibitemShut {NoStop}%
\bibitem [{\citenamefont {Bahadoran}\ and\ \citenamefont
  {Bodineau}(2015)}]{10.1214/14-BJPS277}%
  \BibitemOpen
  \bibfield  {author} {\bibinfo {author} {\bibfnamefont {C.}~\bibnamefont
  {Bahadoran}}\ and\ \bibinfo {author} {\bibfnamefont {T.}~\bibnamefont
  {Bodineau}},\ }\href {\doibase 10.1214/14-BJPS277} {\bibfield  {journal}
  {\bibinfo  {journal} {Brazilian Journal of Probability and Statistics}\
  }\textbf {\bibinfo {volume} {29}},\ \bibinfo {pages} {282 } (\bibinfo {year}
  {2015})}\BibitemShut {NoStop}%
\bibitem [{\citenamefont {Banerjee}\ and\ \citenamefont
  {Basu}(2020)}]{BanerjeeBasu}%
  \BibitemOpen
  \bibfield  {author} {\bibinfo {author} {\bibfnamefont {T.}~\bibnamefont
  {Banerjee}}\ and\ \bibinfo {author} {\bibfnamefont {A.}~\bibnamefont
  {Basu}},\ }\href {\doibase 10.1103/PhysRevResearch.2.013025} {\bibfield
  {journal} {\bibinfo  {journal} {Phys. Rev. Res.}\ }\textbf {\bibinfo {volume}
  {2}},\ \bibinfo {pages} {013025} (\bibinfo {year} {2020})}\BibitemShut
  {NoStop}%
\bibitem [{\citenamefont {Enaud}\ and\ \citenamefont
  {Derrida}(2004)}]{Enaud_2004}%
  \BibitemOpen
  \bibfield  {author} {\bibinfo {author} {\bibfnamefont {C.}~\bibnamefont
  {Enaud}}\ and\ \bibinfo {author} {\bibfnamefont {B.}~\bibnamefont
  {Derrida}},\ }\href {\doibase 10.1209/epl/i2003-10153-8} {\bibfield
  {journal} {\bibinfo  {journal} {Europhys. Lett.}\ }\textbf {\bibinfo {volume}
  {66}},\ \bibinfo {pages} {83} (\bibinfo {year} {2004})}\BibitemShut {NoStop}%
\bibitem [{\citenamefont {Neri}\ \emph {et~al.}(2011)\citenamefont {Neri},
  \citenamefont {Kern},\ and\ \citenamefont
  {Parmeggiani}}]{PhysRevLett.107.068702}%
  \BibitemOpen
  \bibfield  {author} {\bibinfo {author} {\bibfnamefont {I.}~\bibnamefont
  {Neri}}, \bibinfo {author} {\bibfnamefont {N.}~\bibnamefont {Kern}}, \ and\
  \bibinfo {author} {\bibfnamefont {A.}~\bibnamefont {Parmeggiani}},\ }\href
  {\doibase 10.1103/PhysRevLett.107.068702} {\bibfield  {journal} {\bibinfo
  {journal} {Phys. Rev. Lett.}\ }\textbf {\bibinfo {volume} {107}},\ \bibinfo
  {pages} {068702} (\bibinfo {year} {2011})}\BibitemShut {NoStop}%
\bibitem [{\citenamefont {Neri}\ \emph {et~al.}(2013)\citenamefont {Neri},
  \citenamefont {Kern},\ and\ \citenamefont {Parmeggiani}}]{Neri_2013}%
  \BibitemOpen
  \bibfield  {author} {\bibinfo {author} {\bibfnamefont {I.}~\bibnamefont
  {Neri}}, \bibinfo {author} {\bibfnamefont {N.}~\bibnamefont {Kern}}, \ and\
  \bibinfo {author} {\bibfnamefont {A.}~\bibnamefont {Parmeggiani}},\ }\href
  {\doibase 10.1088/1367-2630/15/8/085005} {\bibfield  {journal} {\bibinfo
  {journal} {New Journal of Physics}\ }\textbf {\bibinfo {volume} {15}},\
  \bibinfo {pages} {085005} (\bibinfo {year} {2013})}\BibitemShut {NoStop}%
\bibitem [{\citenamefont {Denisov}\ \emph {et~al.}(2015)\citenamefont
  {Denisov}, \citenamefont {Miedema}, \citenamefont {Nienhuis},\ and\
  \citenamefont {Schall}}]{PhysRevE.92.052714}%
  \BibitemOpen
  \bibfield  {author} {\bibinfo {author} {\bibfnamefont {D.~V.}\ \bibnamefont
  {Denisov}}, \bibinfo {author} {\bibfnamefont {D.~M.}\ \bibnamefont
  {Miedema}}, \bibinfo {author} {\bibfnamefont {B.}~\bibnamefont {Nienhuis}}, \
  and\ \bibinfo {author} {\bibfnamefont {P.}~\bibnamefont {Schall}},\ }\href
  {\doibase 10.1103/PhysRevE.92.052714} {\bibfield  {journal} {\bibinfo
  {journal} {Phys. Rev. E}\ }\textbf {\bibinfo {volume} {92}},\ \bibinfo
  {pages} {052714} (\bibinfo {year} {2015})}\BibitemShut {NoStop}%
\bibitem [{\citenamefont {Concannon}\ and\ \citenamefont
  {Blythe}(2014)}]{ConcannonBlythe}%
  \BibitemOpen
  \bibfield  {author} {\bibinfo {author} {\bibfnamefont {R.~J.}\ \bibnamefont
  {Concannon}}\ and\ \bibinfo {author} {\bibfnamefont {R.~A.}\ \bibnamefont
  {Blythe}},\ }\href {\doibase 10.1103/PhysRevLett.112.050603} {\bibfield
  {journal} {\bibinfo  {journal} {Phys. Rev. Lett.}\ }\textbf {\bibinfo
  {volume} {112}},\ \bibinfo {pages} {050603} (\bibinfo {year}
  {2014})}\BibitemShut {NoStop}%
\bibitem [{\citenamefont {Gran{\'e}li}\ \emph {et~al.}(2006)\citenamefont
  {Gran{\'e}li}, \citenamefont {Yeykal}, \citenamefont {Robertson},\ and\
  \citenamefont {Greene}}]{GraneliGreeneRobertsonYeykal}%
  \BibitemOpen
  \bibfield  {author} {\bibinfo {author} {\bibfnamefont {A.}~\bibnamefont
  {Gran{\'e}li}}, \bibinfo {author} {\bibfnamefont {C.~C.}\ \bibnamefont
  {Yeykal}}, \bibinfo {author} {\bibfnamefont {R.~B.}\ \bibnamefont
  {Robertson}}, \ and\ \bibinfo {author} {\bibfnamefont {E.~C.}\ \bibnamefont
  {Greene}},\ }\href {\doibase 10.1073/pnas.0508366103} {\bibfield  {journal}
  {\bibinfo  {journal} {Proc. Natl. Acad. Sci. U.S.A.}\ }\textbf {\bibinfo
  {volume} {103}},\ \bibinfo {pages} {1221} (\bibinfo {year}
  {2006})}\BibitemShut {NoStop}%
\bibitem [{\citenamefont {Wang}\ \emph {et~al.}(2006)\citenamefont {Wang},
  \citenamefont {Austin},\ and\ \citenamefont {Cox}}]{AustinCoxWang}%
  \BibitemOpen
  \bibfield  {author} {\bibinfo {author} {\bibfnamefont {Y.~M.}\ \bibnamefont
  {Wang}}, \bibinfo {author} {\bibfnamefont {R.~H.}\ \bibnamefont {Austin}}, \
  and\ \bibinfo {author} {\bibfnamefont {E.~C.}\ \bibnamefont {Cox}},\ }\href
  {\doibase 10.1103/PhysRevLett.97.048302} {\bibfield  {journal} {\bibinfo
  {journal} {Phys. Rev. Lett.}\ }\textbf {\bibinfo {volume} {97}},\ \bibinfo
  {pages} {048302} (\bibinfo {year} {2006})}\BibitemShut {NoStop}%
\bibitem [{\citenamefont {Yamamoto}\ \emph {et~al.}(2014)\citenamefont
  {Yamamoto}, \citenamefont {Akimoto}, \citenamefont {Hirano}, \citenamefont
  {Yasui},\ and\ \citenamefont {Yasuoka}}]{AkimotoHiraoYamamotoYasuiYasuoka}%
  \BibitemOpen
  \bibfield  {author} {\bibinfo {author} {\bibfnamefont {E.}~\bibnamefont
  {Yamamoto}}, \bibinfo {author} {\bibfnamefont {T.}~\bibnamefont {Akimoto}},
  \bibinfo {author} {\bibfnamefont {Y.}~\bibnamefont {Hirano}}, \bibinfo
  {author} {\bibfnamefont {M.}~\bibnamefont {Yasui}}, \ and\ \bibinfo {author}
  {\bibfnamefont {K.}~\bibnamefont {Yasuoka}},\ }\href {\doibase
  10.1103/PhysRevE.89.022718} {\bibfield  {journal} {\bibinfo  {journal} {Phys.
  Rev. E}\ }\textbf {\bibinfo {volume} {89}},\ \bibinfo {pages} {022718}
  (\bibinfo {year} {2014})}\BibitemShut {NoStop}%
\bibitem [{\citenamefont {Sakai}\ and\ \citenamefont
  {Akimoto}(2023)}]{sakai2023sample}%
  \BibitemOpen
  \bibfield  {author} {\bibinfo {author} {\bibfnamefont {I.}~\bibnamefont
  {Sakai}}\ and\ \bibinfo {author} {\bibfnamefont {T.}~\bibnamefont
  {Akimoto}},\ }\href@noop {} {\bibfield  {journal} {\bibinfo  {journal}
  {arXiv:2301.00563}\ } (\bibinfo {year} {2023})}\BibitemShut {NoStop}%
\bibitem [{sup()}]{supplementary}%
  \BibitemOpen
  \href@noop {} {\bibinfo  {journal} {See Supplementary Material for the
  passage time}\ }\BibitemShut {NoStop}%
\bibitem [{\citenamefont {Godr{\`e}che}\ and\ \citenamefont
  {Luck}(2001)}]{Godreche}%
  \BibitemOpen
\bibfield  {journal} {  }\bibfield  {author} {\bibinfo {author} {\bibfnamefont
  {C.}~\bibnamefont {Godr{\`e}che}}\ and\ \bibinfo {author} {\bibfnamefont
  {J.~M.}\ \bibnamefont {Luck}},\ }\href {\doibase 10.1023/A:1010364003250}
  {\bibfield  {journal} {\bibinfo  {journal} {J. Stat. Phys.}\ }\textbf
  {\bibinfo {volume} {104}},\ \bibinfo {pages} {489} (\bibinfo {year}
  {2001})}\BibitemShut {NoStop}%
\bibitem [{\citenamefont {Feller}(1971)}]{Feller1971}%
  \BibitemOpen
  \bibfield  {author} {\bibinfo {author} {\bibfnamefont {W.}~\bibnamefont
  {Feller}},\ }\href@noop {} {\emph {\bibinfo {title} {An Introduction to
  Probability Theory and its Applications}}},\ \bibinfo {edition} {2nd}\ ed.,\
  Vol.~\bibinfo {volume} {2}\ (\bibinfo  {publisher} {Wiley, New York},\
  \bibinfo {year} {1971})\BibitemShut {NoStop}%
\bibitem [{\citenamefont {de~Haan}\ and\ \citenamefont
  {Ferreira}(2006)}]{HaanFerreira}%
  \BibitemOpen
  \bibfield  {author} {\bibinfo {author} {\bibfnamefont {L.}~\bibnamefont
  {de~Haan}}\ and\ \bibinfo {author} {\bibfnamefont {A.}~\bibnamefont
  {Ferreira}},\ }\href@noop {} {\emph {\bibinfo {title} {Extreme value theory:
  an introduction}}},\ Vol.~\bibinfo {volume} {21}\ (\bibinfo  {publisher}
  {Springer},\ \bibinfo {year} {2006})\BibitemShut {NoStop}%
\end{thebibliography}%

\end{document}

%section {title}
%\preprint{APS/123-QED}

\title{Supplementary Material for ``Non-self-averaging of current\\in a totally asymmetric simple exclusion process with quenched disorders''}% Force line breaks with \\

\author{Issei Sakai}
% \email{6222514@ed.tus.ac.jp}
\affiliation{%
  Department of Physics, Tokyo University of Science, Noda, Chiba 278-8510, Japan
}%

% \author{Keiji Saito}
% \affiliation{%
%   Department of Physics, Keio University, Yokohama, 223-8522, Japan
% }%

\author{Takuma Akimoto}
% \email{takuma@rs.tus.ac.jp}
\affiliation{%
  Department of Physics, Tokyo University of Science, Noda, Chiba 278-8510, Japan
}%

%}


\date{\today}% It is always \today, today,

%\pacs{05.45.Ac, 05.40.Fb, 87.15.Vv}% PACS, the Physics and Astronomy
\maketitle

%\tableofcontents
\section{Passage time}
\begin{figure}[b]
    \centering
    \includegraphics[width=8.6cm]{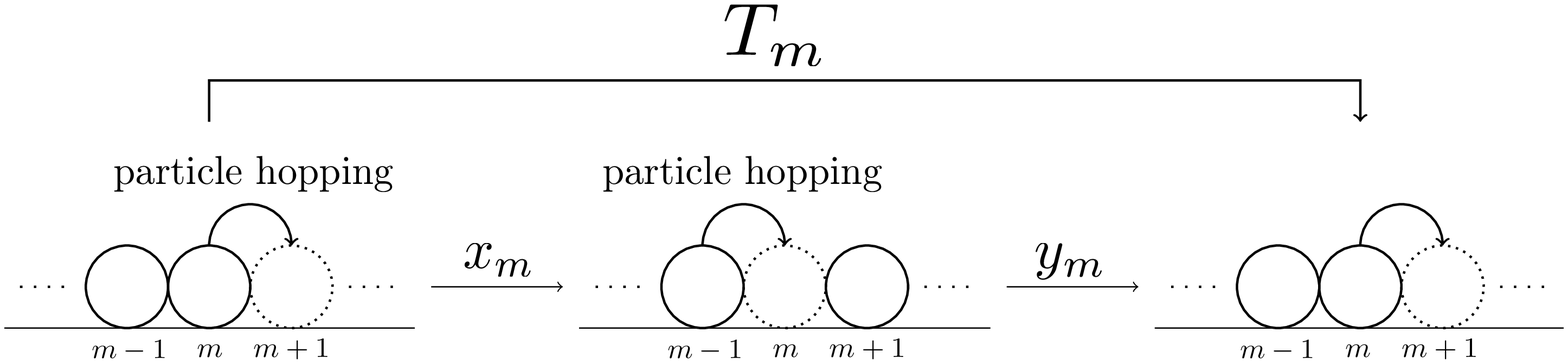}
    \caption{Particle dynamics during the passage time.
    The solid- and dotted-line circles denote particles and holes, respectively.}
    \label{fig7}
\end{figure}

Here, we derive the distribution of the passage time $T_m$ between sites $m$ and $m+1$ when the current is maximized.
We note that the waiting time at site $m$ is maximum.
The passage time  can be divided into the hole-escape $x_m$ and particle-escape time $y_m$.
The hole- or particle-escape times are defined as the time until the hole or particle hops from site $m$,
respectively (Fig.~\ref{fig7}).
At site $m$, when a hole attempts to hop $i$ times, the probability density function (PDF) of the waiting time to hop to the right site follows the Erlang distribution,
\begin{equation}
    f_i(x_m)=\frac{x_m^{i-1}}{(i-1)!\tau_{m-1}^i}\exp{\left(-\frac{x_m}{\tau_{m-1}}\right)},
\end{equation}
and the probability of completing the hopping process by attempting $i$th times, i.e., $(i-1)$th failures and success at the $i$th attempt,
is $\rho_{m-1}(1-\rho_{m-1})^{i-1}$.
Thus, the PDF $f_{\mathrm{HD}}(x_m)$ of $x_m$ is given by
\begin{equation}
    \begin{split}
        f_{\mathrm{HD}}(x_m)&=\rho_{m-1}\sum_{i=1}^\infty (1-\rho_{m-1})^{i-1}f_i(x_m)\\
        &=\frac{\rho_{m-1}}{\tau_{m-1}}\exp{\left(-\frac{x_m}{\tau_{m-1}}\right)}\sum_{i=1}^\infty\left(
            \frac{1-\rho_{m-1}}{\tau_{m-1}}x_m
        \right)^{i-1}\\
        &=E\left(x_m;\frac{\tau_{m-1}}{\rho_{m-1}}\right),
    \end{split}
    \label{hole-escape}
\end{equation}
where $E(x;\theta)\equiv \exp{(-x/\theta)}/\theta$ is the exponential distribution and $\theta$ is the mean value.
At site $m+1$, using the same method of deriving Eq.~(\ref{hole-escape}), the PDF $g_{\mathrm{LD}}(y_{m+1})$ of $y_{m+1}$ can be obtained as
\begin{equation}
    g_{\mathrm{LD}}(y_{m+1})=E\left(y_{m+1};\frac{\tau_{m+1}}{1-\rho_{m+2}}\right).
    \label{particle-escape_LD}
\end{equation}
Using Eq.~(\ref{particle-escape_LD}), we can derive the joint PDF of $x_m$ and $y_m$.
When a particle at site $m$ completes hopping by attempting once, $y_{m+1}<x_m+y_m$.
Then, the weighted joint PDF $h_1(x_m,y_m)$ of $x_m$ and $y_m$ is given by
\begin{equation}
    \begin{split}
        h_1(x_m,y_m)=&f_{\mathrm{HD}}(x_m)E(y_m;\tau_m)\\
        &\times\int_0^{x_m+y_m}dy_{m+1}g_{\mathrm{LD}}(y_{m+1}).
    \end{split}
\end{equation}
When a particle at site $m$ completes as hopping by attempting $i$th times ($i>1$), $x_m+y_m'<y_{m+1}<x_m+y_m$,
where $y_m'$ is the waiting time to attempt hopping for $(i-1)$th times.
The PDF of $y_m'$ follows the Erlang distribution,
\begin{equation}
    g_{i-1}(y_m')=\frac{(y_m')^{i-2}}{(i-2)!\tau_m^{i-1}}\exp{\left(-\frac{y_m'}{\tau_m}\right)}.
\end{equation}
Then, the weighted joint PDF $h_i(x_m,y_m)$ of $x_m$ and $y_m$ is given by

\begin{figure}[b]
    \centering
    \includegraphics[width=8.6cm]{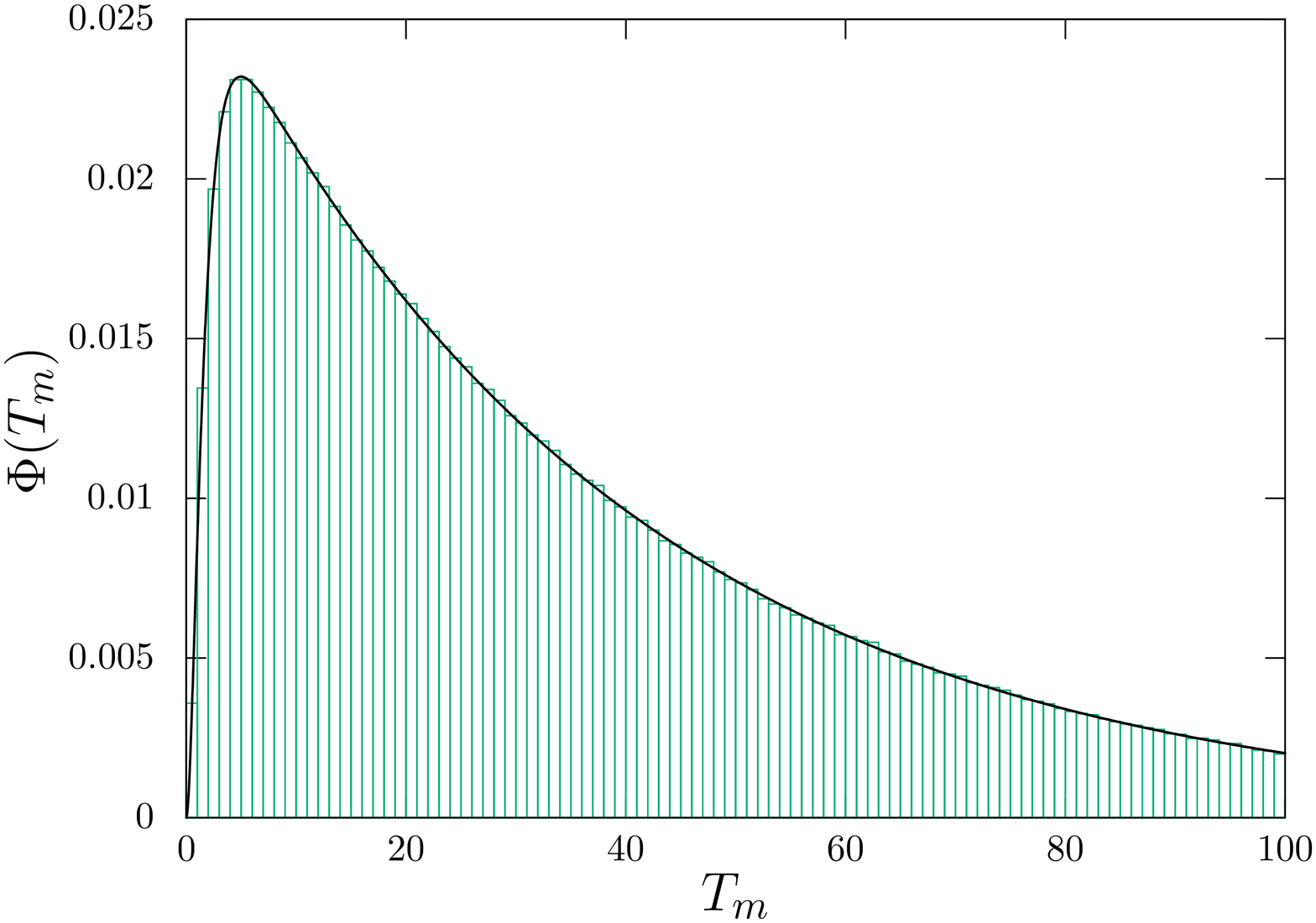}
    \caption{Distribution of the passage times between site $m$ and $m+1$. 
    The bars are histograms obtained from the numerical simulation ($\alpha=2.5$ and $L=5000$). 
    The solid line represents Eq.~(\ref{passage-time}).}
    \label{fig8}
\end{figure}
\begin{widetext}
\begin{equation}
    h_i(x_m,y_m)=\int_0^{y_m}dy_m'~g_{i-1}(y_m')E(y_m-y_m';\tau_m)\int_{x_m+y_m'}^{x_m+y_m}dy_{m+1}~g_{\mathrm{LD}}(y_{m+1}).
\end{equation}
Therefore, the joint PDF $h(x_m,y_m)$ of $x_m$ and $y_m$ is given by
\begin{equation}
    \begin{split}
        h(x_m,y_m)=&\sum_{i=1}^\infty h_i(x_m,y_m)\\
        =&f_{\mathrm{HD}}(x_m)E(y_m;\tau_m)+\frac{1}{\tau_m\frac{1-\rho_{m+2}}{\tau_{m+1}}-1}
        \exp{\left(-\frac{1-\rho_{m+2}}{\tau_{m+1}}x_m\right)}f_{\mathrm{HD}}(x_m)
        (E(y_m;\tau_m)-g_{\mathrm{LD}}(y_m)).
    \end{split}
\end{equation}
Thus, the PDF $\Phi(T_m)$ of $T_m$ is given by
\begin{equation}
    \begin{split}
        \Phi(T_m)=&\int_0^{T_m} dx~h(x,T_m-x)\\
        =&\tau_m\frac{\rho_{m-1}}{\tau_{m-1}}\left(\zeta_1+\zeta_2\zeta_3\right)E(T_m;\tau_m)-\zeta_1f_{\mathrm{HD}}(T_m)-\zeta_2g_{\mathrm{LD}}(T_m)+\zeta_3E\left(T_m;\frac{1}{\frac{\rho_{m-1}}{\tau_{m-1}}+\frac{1-\rho_{m+2}}{\tau_{m+1}}}\right),
    \end{split}
    \label{passage-time}
\end{equation}
where
\begin{align}
    \zeta_1\equiv \frac{1}{\tau_m\frac{\rho_{m-1}}{\tau_{m-1}}-1},\ 
    \zeta_2\equiv \frac{1}{\tau_m\frac{1-\rho_{m+2}}{\tau_{m+1}}-1},\ 
    \zeta_3\equiv \frac{1}{\tau_m\left(\frac{\rho_{m-1}}{\tau_{m-1}}+\frac{1-\rho_{m+2}}{\tau_{m+1}}\right)-1}.
\end{align}

Next, we derive the mean and variance of $T_m$.
The Laplace transform of $\Phi(T_m)$ with respect to $s$ is given by
\begin{align}
    \hat{\Phi}(s)&\equiv \int_0^\infty dT_m e^{-sT_m}\Phi(T_m)\notag\\
    &=\tau_m\frac{\rho_{m-1}}{\tau_{m-1}}\left(\zeta_1+\zeta_2\zeta_3\right)\frac{1}{\tau_m s+1}-\frac{\zeta_1}{\frac{\tau_{m-1}}{\rho_{m-1}}s+1}-\frac{\zeta_2}{\frac{\tau_{m+1}}{1-\rho_{m+2}}+1}+\frac{\zeta_3}{\frac{s}{\frac{\rho_{m-1}}{\tau_{m-1}}+\frac{1-\rho_{m+2}}{\tau_{m+1}}}+1}
\end{align}
Thus, the mean and variance of the passage time are given by
\begin{align}
    &\braket{T_m}=\tau_m+\frac{\tau_{m-1}}{\rho_{m-1}}+\frac{\frac{\rho_{m-1}}{\tau_{m-1}}}{\frac{\rho_{m-1}}{\tau_{m-1}}+\frac{1-\rho_{m+2}}{\tau_{m+1}}}\frac{\tau_{m+1}}{1-\rho_{m+2}},\\
    &\braket{T_m^2}-\braket{T_m}^2=\tau_m^2+\left(\frac{\tau_{m-1}}{\rho_{m-1}}\right)^2+\left(\frac{\tau_{m+1}}{1-\rho_{m+2}}\right)^2
    -\frac{3}{\left(\frac{\rho_{m-1}}{\tau_{m-1}}+\frac{1-\rho_{m+2}}{\tau_{m+1}}\right)^2}.
\end{align}
\end{widetext}